\documentclass{aa}  

\usepackage{ulem}
\renewcommand{\arraystretch}{1.1}
\usepackage{color}
\usepackage{array}
\usepackage{url}
\usepackage[bottom]{footmisc}
\usepackage{adjustbox}
\usepackage{comment}

\newcommand{\nustar}{{\sl NuSTAR}\xspace}
\newcommand{\nicer}{{\rm NICER}\xspace}
\newcommand{\ngc}{{NGC 4190 ULX-1}\xspace}

\usepackage{graphicx}
\usepackage{caption}
\usepackage{subcaption}
\usepackage{booktabs}
\usepackage{adjustbox}
\usepackage[toc,page]{appendix}
\usepackage{amsmath}
\usepackage{nccmath}
\usepackage{txfonts}
\usepackage[]{hyperref}
\hypersetup{unicode=true, colorlinks=true, linkcolor=[rgb]{0.53, 0.15, 0.34}, citecolor=[rgb]{0.0, 0.27, 0.42}, filecolor=[rgb]{1.0, 0.13, 0.32}, urlcolor=[rgb]{0.53, 0.15, 0.34}}
\usepackage{ulem}

\begin{document}

   \title{Simultaneous \nicer and \nustar observations of the ultraluminous source NGC 4190 ULX-1}

  \subtitle{}

   \author{Jorge A. Combi\inst{1,2,5}, Federico A. Fogantini\inst{1}, Enzo A. Saavedra\inst{3,4}, Gustavo E. Romero\inst{1,2}, Leandro Abaroa\inst{1,2}, Federico Garc\'{\i}a\inst{1,2}, Pedro Luque-Escamilla\inst{5}, Josep Martí\inst{5}, Nelson Cruz-Sanchez\inst{2}
     }

   \institute{Instituto Argentino de Radioastronom\'ia (CCT La Plata, CONICET; CICPBA; UNLP), C.C.5, (1894) Villa Elisa, Argentina \and Facultad de Ciencias Astron\'omicas y Geof\'{\i}sicas, Universidad Nacional de La Plata, B1900FWA La Plata, Argentina \and
   Instituto de Astrofísica de Canarias (IAC), Vía Láctea s/n, La Laguna 38205, S/C de Tenerife, Spain \and
   Departamento de Astrofísica, Universidad de La Laguna, La Laguna, E-38205, S/C de Tenerife, Spain \and
   Departamento de F\'isica (EPS), Universidad de Ja\'en, Campus Las Lagunillas s/n, A3, 23071 Ja\'en, Spain}

   \date{Received; accepted}

 
  \abstract
   {}
   {We present an X-ray analysis of three different {\sl XMM-Newton} observations together with simultaneous \nicer and \nustar observations of the ultraluminous X-ray source NGC 4190 ULX-1. Our goal is to constrain the structure of the accretion disk and the geometrical properties of the source.}
   {We performed temporal and spectral analyses in the 0.4--30 keV energy range in which the source is significantly detected in dedicated {\sl XMM-Newton}, \nicer, and {\sl NuSTAR} observations.}
   {The temporal analysis shows no flaring activity in the light curves. No pulsation is detected throughout. The source exhibits a typical ULX spectrum, which can be fitted with two thermal blackbody components plus a Comptonization tail at high energies. The luminosity-temperature ($L-T$) relation of each thermal spectral component is consistent with the $L \propto T^{2}$ relation expected from an advection-dominated supercritical disk.}
   {We interpret these results as a super-Eddington accreting black hole seen almost face-on.  A dense wind ejected from the disk obscures the central source, and a hot electron plasma is evacuated through the funnel formed above the hole. Geometric beaming is responsible for the ULX soft emission, whereas the hard tail is the result of the Comptonization of soft photons by the electrons ejected through the funnel.}

   \titlerunning{Simultaneous \nicer and {\sl NuSTAR} observations of NGC 4190 ULX-1}
   \authorrunning{Jorge A. Combi et al.}
   \maketitle
%

\section{Introduction}

Ultraluminous X-ray sources (ULXs) provide one of the best opportunities to study super-Eddington accretion in nearby galaxies \citep{2017ARA&A..55..303K}. In the last decade, the nature of these off-nuclear point sources with luminosities $\geq$ 10$^{39}$ erg s$^{-1}$ has been the subject of considerable debate. Several hypotheses have been proposed to explain the high X-ray luminosities. These include i) intermediate-mass black holes (BHs) accreting at sub-Eddington rates \citep{1999ApJ...519...89C}, ii) sub-Eddington accreting compact stellar-mass objects with jets \citep{2006MNRAS.370..399B}, iii) stellar-mass BHs accreting at supercritical rates \citep{2007MNRAS.377.1187P,2021AstBu..76....6F}, and iv) neutron stars (NSs) accreting at supercritical rates \citep{Mushtukov2017MNRAS.467.1202M, Koliopanos2017A&A...608A..47K}.

Ultraluminous X-ray sources are classified into different states based on their X-ray spectral components and hardness \citep{Sutton2013MNRAS,2017ARA&A..55..303K}. The most common state is the ultraluminous state, which is characterized by a power-law tail with a cutoff \citep{Gladstone2009MNRAS.397.1836G}. Ultraluminous X-ray sources in this state can be further classified as hard ultraluminous (HUL) or soft ultraluminous (SUL) sources depending on their hardness \citep{Sutton2013MNRAS}. Soft bright states are generally in a higher luminosity range compared to hard intermediate states, due to the down-scattering of photons by the dense medium of clumpy winds at high accretion rates. Other ULX states are the broadened disk (BD) state and the super-soft ultraluminous (SSUL) state. Broadened disk states are characterized by  a pure accretion disk model with a temperature of 1--2.5 keV, while SSUL states are dominated by single-component cool blackbody emissions. It should be noted that the distinction between the above classifications is not always clear. A significant number of ULXs exhibit spectral transitions between these different states or regimes \citep[e.g.][]{Gurpide2021A&Ab}. These changes are likely due to variations in the accretion rate, disk occultation, wind outflow strength, precession of the accretion disk, and various other physical parameters.
In recent years, observations performed with a new generation of X-ray telescopes like {\sl XMM-Newton}, {\sl Chandra}, \nicer, or {\sl NuSTAR} have allowed us to establish the emission mechanism for the majority of ULX sources as super-Eddington accretion from a stellar mass X-ray binary system \citep{2013ApJ...779..148W,2015ApJ...808...64M}. This result was further supported by the remarkable discovery that several ULXs are actually powered by accreting NSs \citep{2014Natur.514..202B}.

\begin{figure*}
\centering
   \includegraphics[width=0.99\textwidth]{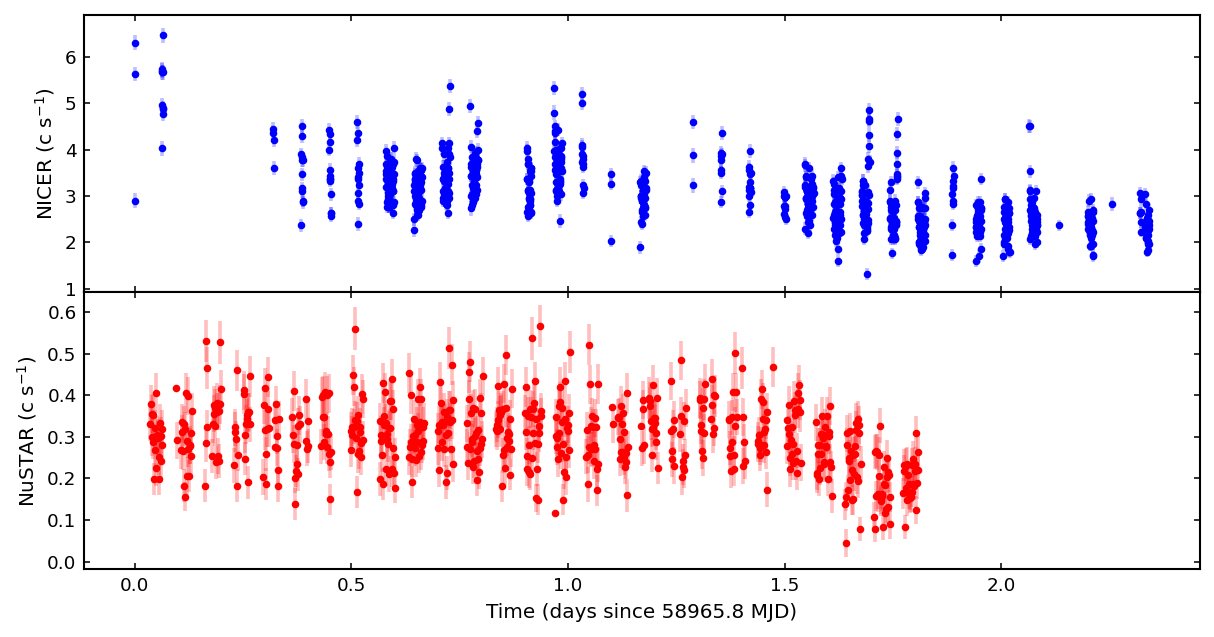}
   \caption{\nicer (0.3-12 keV, top panel) and \nustar (3-79 keV, bottom panel) background substracted light-curves of NGC 4190 ULX-1 with a time bin of 30~s and 120~s respectively.}
   \label{Fig:lc}
\end{figure*}

The first physical models of super-Eddington accretion involved a geometrically thin and optically thick disk around the central compact object \citep{1973A&A....24..337S}. One of the main predictions of these models is massive radiation-driven outflows of ionized gas from regions close to the compact object, escaping at slightly relativistic velocities. However, most current models consider disk inflation inside the critical radius, where the radiation pressure exceeds the gravitational attraction. In this inner region, the disk is unstable and most of the matter is evacuated by a very dense, opaque wind. Photon trapping and advection remove energy from the disk, feeding the BH at a rate close to the Eddington accretion rate \citep{2003ApJ...596..429O,2004PASJ...56..569F,2011MNRAS.413.1623D,2017PASJ...69...33O}.

In this context, of particular interest is the study of super-Eddington accretion in the ULX system NGC 4190 ULX1, located at a distance, $D \approx$ 2.9~Mpc \citep{2013AJ....146...86T}. This source was recently studied by \citet{Ghosh2021}. Of the detailed analysis of X-ray spectra from several {\sl XMM-Newton} observations, the authors conclude that ULX1 in NGC 4190 it is not in a standard canonical accretion state, but in an ultraluminous regime. In addition, the positive luminosity-temperature relation suggests that the multicolor disk model seems to favor the $L\propto$ T$^2$ relation, consistent with an advection-dominated inflated disk emission. 

Recently, \cite{2023A&A...671A...9A} implemented a physical model that fits the available {\sl XMM-Newton} data for this source. The model is based on a BH of $M_{\rm BH}=10\; M_{\odot}$ accreting at $\dot{M}\sim 10 \;\dot{M}_{\rm Edd}$ (where $\dot{M}_{\rm Edd}\approx2.2\times 10^{-8}M_{\rm BH}\,{\rm yr^{-1}}$ is the Eddington rate). The system is viewed almost face-on, with an inclination of $i\approx 0^{\circ}$.

In this paper we present an analysis of archival {\sl XMM-Newton}, {\rm NICER}, and {\sl NuSTAR} observations of the ULX source NGC 4190 ULX-1 to better constrain the geometry and structure of the accretion disk around the BH, to detect spectral signatures of outflows, and to search for possible pulsations. 

The paper is organized as follows. In \hyperref[sec:data]{section~\ref{sec:data}} we provide details about the observations used in this work. In Sect. 3 we describe the main results of the temporal and spectral analyses, while in Sect. 4 we discuss the behavior of the thermal components and present an improved model of the source. Finally, in Sect. 5 we present our conclusions.

\begin{figure*}
\centering
   \includegraphics[width=0.99\columnwidth]{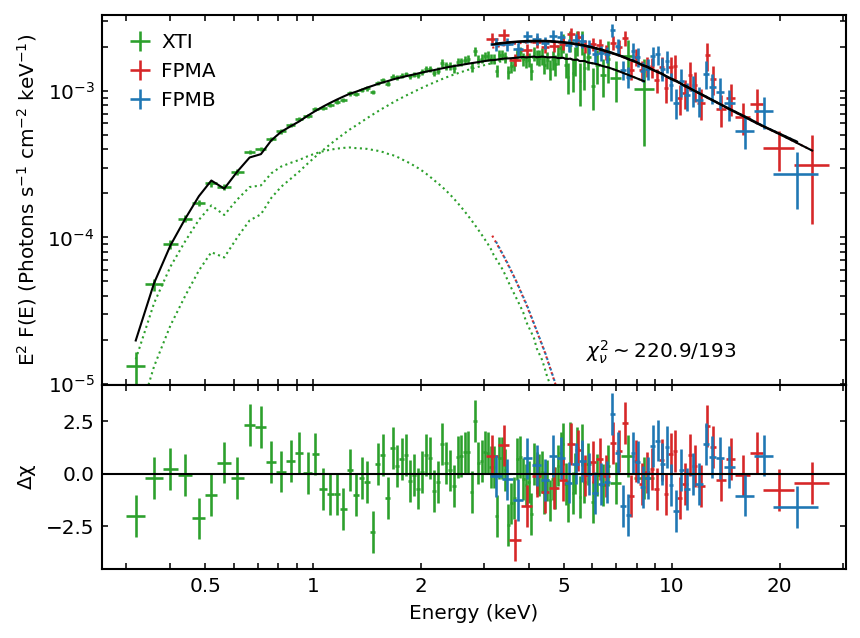}
    \includegraphics[width=0.99\columnwidth]{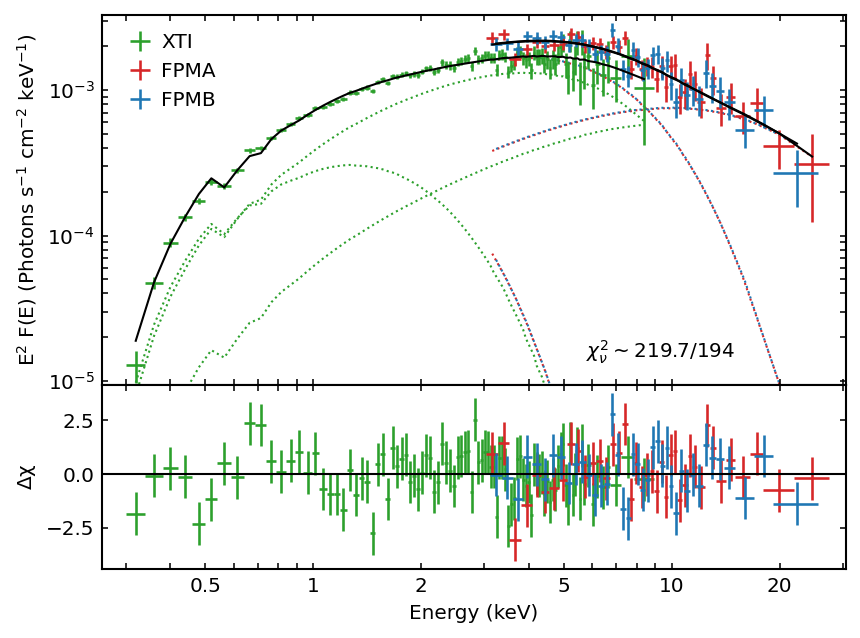}
   \caption{\nicer+\nustar unfolded spectrum and residuals from both tested models. The \nicer background dominated above 9 keV, while the \nustar background dominated above 30 keV. {\sl Left panel}: {\sc simpl} model. {\sl Right panel}: {\sc cutoffpl} model.}
   \label{fig:datafit}
\end{figure*}

\section{Observation and data analysis} \label{sec:data}

\subsection{{\sl XMM-Newton} data}

{\sl XMM-Newton} observed NGC 4190 ULX1 on May 6, 2010 with an exposure time of 21 ks (ObsID 0654650101), on May 8, 2010 with an exposure time of 27 ks (ObsID 0654650201), and November 25, 2010 with an exposure time of 20 ks (ObsID 0654650301). In the three observations, the MOS and pn cameras were in full-frame mode.

The {\sl XMM-Newton} data were reduced using v21.0.0 of the {\sl XMM-Newton} Science Analysis System (SAS) software and up-to-date CCFs as of October 2023, producing calibrated event files with {\tt epproc} and {\tt emproc} and removing periods of high background. We extracted data products using circular regions of 30 and 60 arcsec radius for the source and background, respectively, with background regions located on the same chip at a similar distance from the readout node. We selected events with {\tt FLAG == 0}, with {\tt PATTERN <= 12} for the EPIC-MOS cameras and {\tt PATTERN <= 4} for the EPIC-pn cameras. In all cases, spectra were grouped into having a minimum of 20 counts per bin. Redistribution matrices and auxiliary response files were generated with the tasks {\tt rmfgen} and {\tt armgen}, respectively. 

\subsection{\nicer data}

The Neutron star Interior Composition ExploreR (\nicer; \citealt{2016GendreauSPIE.9905E..1HG}) is an instrument on board the International Space Station equipped with an X-ray timing instrument (XTI) detector.  The X-ray timing instrument detects soft X-rays in the 0.2--12 keV energy range with a high temporal resolution. We analyzed four \nicer's ObsIDs 3645010101/2/3/4 performed between April 26, 2020 and April 28, 2020, with exposures of $\sim$ 0.7, 13, 17, and 2.9~ks, respectively. We followed the standard data processing using {\tt HEASoft}~v.6.32.1 and \nicer data analysis software ({\tt xti20221001}). We extracted light curves and spectra using the {\tt nicerl3} task. In addition, for each observation, we obtained the background light curve and spectrum with the {\tt Space Weather} background model \citep{2022AJ....163..130R}. We applied a barycentric correction using the {\tt barycorr} tool.

\subsection{{\sl NuSTAR} Data}

The {\sl NuSTAR} telescope \citep[{\sl Nuclear Spectroscopic Telescope Array};][]{2013ApJ...770..103H} is an X-ray satellite equipped with two detectors, FPMA and FPMB, operating in the 3 to 78 keV energy band. \nustar observed NGC 4190 ULX-1 in April 2020 (ObsID 30601009002), with an exposure of $\sim$ 73 ks. We reduced the data using the analysis software {\tt NuSTARDAS-v. 2.0.0} from {\tt HEASoft}~v.6.32.1 and {\tt CALDB} (V.20130509) calibration files. We took source events that accumulated within a circular region of a 60~arcsec radius around the focal point. The chosen radius encloses ${\sim}75$\% of the point spread function (PSF). Background regions for FPMA/B modules were selected far from contamination of the source and detector edges, with a radius of 100 arcsec.

We used the {\tt nupipeline} task to create level 2 data products, with {\tt saacalc=1}, {\tt saamode=OPTIMIZED}, and {\tt tentacle=NO} to filter high background epochs, obtained from the South Atlantic Anomaly (SAA)
filtering report. This led to us losing $\sim$~0.6~\% of the observation total exposure. We extracted the light curves and spectra with the {\tt nuproducts} task. We obtained the barycenter-corrected light curves using the {\tt barycorr} task with a {\tt nuCclock20100101v136} clock correction file.  We used celestial coordinates $\alpha=183.434049$ deg and $\delta=+36.631493$~deg for the barycentric correction. For this, we used {\tt JPL-DE200} as the Solar System ephemeris. We finally added the light curves of both detectors with the task {\tt LCMATH}. 

\section{Results} \label{sec:results}


\subsection{Light curve analysis} \label{sec:lc}

\hyperref[Fig:lc]{Figure~\ref{Fig:lc}} shows the background-corrected barycentered light curves observed simultaneously by \nicer and \nustar over a time interval of ${\sim}$50 hours. We did not observe any flaring activity in the light curve. In both light curves there is a slight decrease in the count rate toward the end of the observation. To investigate this phenomenon, we obtained light curves in several energy bands with both instruments (0.3--3 keV, 3--12 keV, 12--79 keV) and found that such a decrease in counts is observed up to $\sim$ 12 keV. Above 12 keV, the light curves show a constant behavior, with an average rate of $\sim$ 0.15 c/s. We presume that the observed soft flux decrease and spectral change can be attributed to changes in local absorption.


We used spectral timing routines from the {\tt HENDRICS} \citep{hendrics} and {\tt Stingray} \citep{stingray2019ApJ...881...39H} packages to search for possible pulsations on {\rm NICER} and \nustar data within 0.01--10 Hz frequency interval. We visually inspected power spectra of both instruments and found no pulsations that were significantly above noise. We applied the {\sc HENaccelsearch} tool, which searches for the best pulsation candidates within a given frequency interval, and with varying period derivatives up to $10^{-9}$~Hz~s$^{-1}$. The number of harmonics in the $Z^2_n$ statistic was limited to two. No pulsation candidates were found using the above criteria.

In order to derive an upper limit on the pulsed fraction (PF) that any signal could have, we applied the {\sc HENz2vspf} tool. This task simulates N datasets with the same good time intervals (GTIs) and number of events as the real data, and produces a $Z^2_2$ versus PF for each simulated dataset. Thus, for $N=10000$ simulations we derive an upper limit on the PF (within 90\% confidence) of ${\lesssim}7$\% for \nicer data (0.3--12 keV), and of ${\lesssim}18$\% for {\sl NuSTAR} (3--79 keV, FPMA+B) data.

\subsection{Spectral analysis} \label{sec:spectral}

\renewcommand{\arraystretch}{1.5} 
\begin{table}
\begin{center}

\caption{Best fit parameters.}
\label{tab:params}
\begin{adjustbox}{max width=\columnwidth}
\begin{tabular}{ l  c  c  c}
    \hline
    Component           & Parameter                             & \multicolumn{2}{c}{Model} \\ 
                        &                                       & {\sc simpl}   & {\sc cutoffpl} \\\hline
    {\sc cons}          & $C_{\rm A/XTI}$                       & $1.26\pm0.06$                          & $1.26\pm0.06$               \\
                        & $C_{\rm B/XTI}$                       & $1.29\pm0.06$                          & $1.28\pm0.06$           \\ 
    {\sc tbabs}         & $N_{\rm H}$ (10$^{22}$~cm$^{-2}$)     & $0.09_{-0.01}^{+0.02}$                 & $0.09\pm0.02$           \\ 
    {\sc diskbb}        & $kT_{\rm in}$ (keV)                   & $0.4\pm0.1$                            & $0.4\pm0.1$ \\                  
    {\sc cutoffpl}      & $\Gamma$                              & -                                      & $0.59^{\dagger}$         \\
                        & $E_{\rm fold}$ (keV)                  & -                                      & $7.1^{\dagger}$         \\ 
    {\sc simpl}         & $\Gamma$                              & $3.4\pm0.3$                            & -                       \\
                        & $CF$                                  & $0.9^{+0.1}_{-0.4}$                    & -                       \\ 
    {\sc diskpbb}       & $kT$ (keV)                            & $1.2^{+0.3}_{-0.1}$                    & $1.7_{-0.1}^{+0.2}$             \\ 
                        & $p$                                   & $0.8_{-0.1}^{+0.2}$                    & $0.7_{-0.1}^{+0.2}$  \\                     
    
    Luminosity          & $10^{39}$~erg~s$^{-1}$                & \multicolumn{2}{c}{$7.6\pm0.2$}        \\ \hline 
    $\chi^2$/dof        &                                       & 220.9 / 193                               & 219.7 / 194  \\ \hline 
    
\end{tabular}
\end{adjustbox}
\end{center}
\footnotesize{\textbf{Notes.} Best fit parameters and 90\% level uncertainties derived from NICER+\nustar average spectra. The 0.3--30 keV unabsorbed luminosity was calculated adopting a distance of 2.9~Mpc \citep{2013AJ....146...86T}. The $\dagger$ symbol indicates that the parameters was frozen during the fit.}
\end{table}

We performed the X-ray spectral analysis using {\tt XSPEC} v.12.13.0c~\citep{1996ASPC..101...17A}. The spectra were grouped with the optimal binning method from \cite{Kaastra2016} with a minimum of 20 counts per bin to properly use the $\chi^2$ statistics. Unabsorbed luminosities were estimated using the {\sc cglumin} convolution model, assuming a distance of 2.9 Mpc \citep{2013AJ....146...86T}.

A broadband spectral analysis of the data was performed to characterize the source spectrum. The source were modeled simultaneously using all three detectors (FPMA, FPMB, and XTI). To account for the systematic differences among the three detectors, a calibration constant was assigned to each one. For the \nicer spectra, we noticed channels between 0.3 and 9 keV where the source was above the background. Similarly, for the {\sl NuSTAR} spectrum, we noticed channels between 3 and 30 keV. Consequently, we modeled the entire spectra between 0.3 and 30 keV.

To investigate the properties of \ngc, we have constructed two tentative scenarios. First, we hypothesize a non-magnetic accretor, such as a BH. Second, we consider a magnetic accretor, such as a ULX pulsar, although no pulsations have been detected in \ngc so far. Two neutral absorption models were included in this study, where we fixed the first model to a galactic absorption column of $2.4\times10^{20}$ cm$^{-2}$ derived from the NH tool\footnote{\url{https://heasarc.gsfc.nasa.gov/cgi-bin/Tools/w3nh/w3nh.pl}} and left the second model free to account for local absorption. For the neutral absorption we used the abundances from \citet{2000ApJ...542..914W} and the effective cross sections from \citet{1996ApJ...465..487V}.


Two thermal components were used to represent the accretion flow scenario for non-magnetic objects. The {\sc diskbb} model \citep{Mitsuda1984PASJ...36..741M} was adopted to describe the flow in the outer region or wind with significant opacity, while the {\sc diskpbb} model \citep{Mineshige1994ApJ...435L.125M} was implemented to describe the flow in the inner region where it is expected to deviate from the thin disk approximation. The {\sc diskbb} model assumes a thin disk profile with a fixed exponent $p$ value of 0.75. Conversely, the {\sc diskpbb} model allows $p$ to be set as a free parameter. This combination of models is often implemented to fit low-energy X-ray data ($E < 10$~keV) when analyzing the spectrum of ULXs \citep{Stobbart2006MNRAS,Walton2014ApJ,Walton2017ApJ,Rana2015ApJ,Mukherjee2015ApJ}.

The high-energy emission ($E>10$~keV) is probably associated with Compton scattering of photons from the disk in a hot electron plasma. This process is similar to what occurs in X-ray binaries with BHs below the Eddington limit and in active galactic nuclei \citep{Haardt1991ApJ...380L..51H}. Therefore, we modeled the emission by including an extra high-energy power-law component. To avoid incorrect extrapolations of this power-law emission at extremely low energies, we used the {\sc simpl} model \citep{Steiner2009PASP..121.1279S}. This additional component complements the {\sc diskpbb} component, which is the hotter of the two with respect to the disk. 

\begin{figure*}
\centering
   \includegraphics[width=\columnwidth]{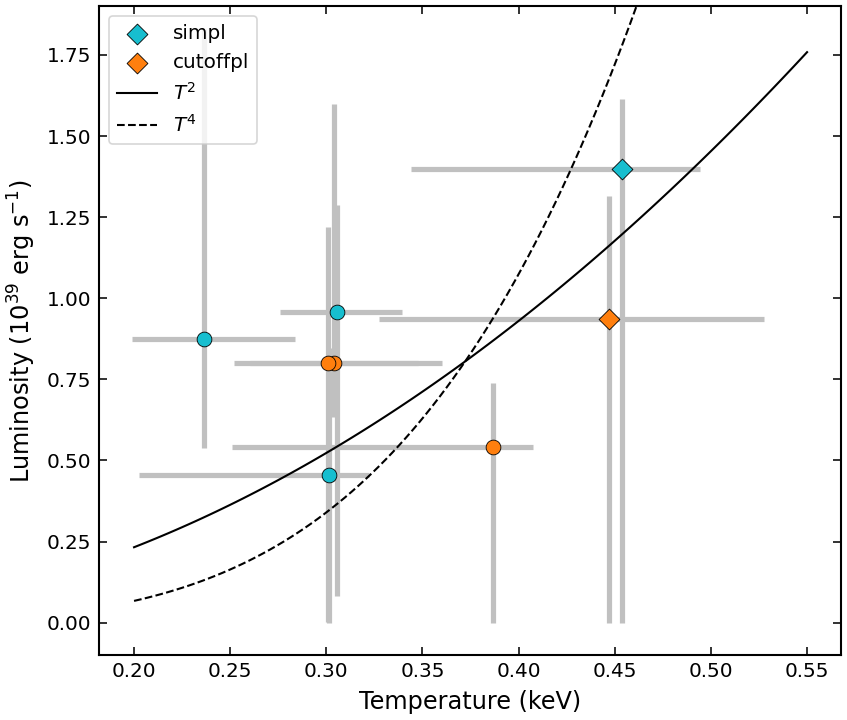}
   \includegraphics[width=\columnwidth]{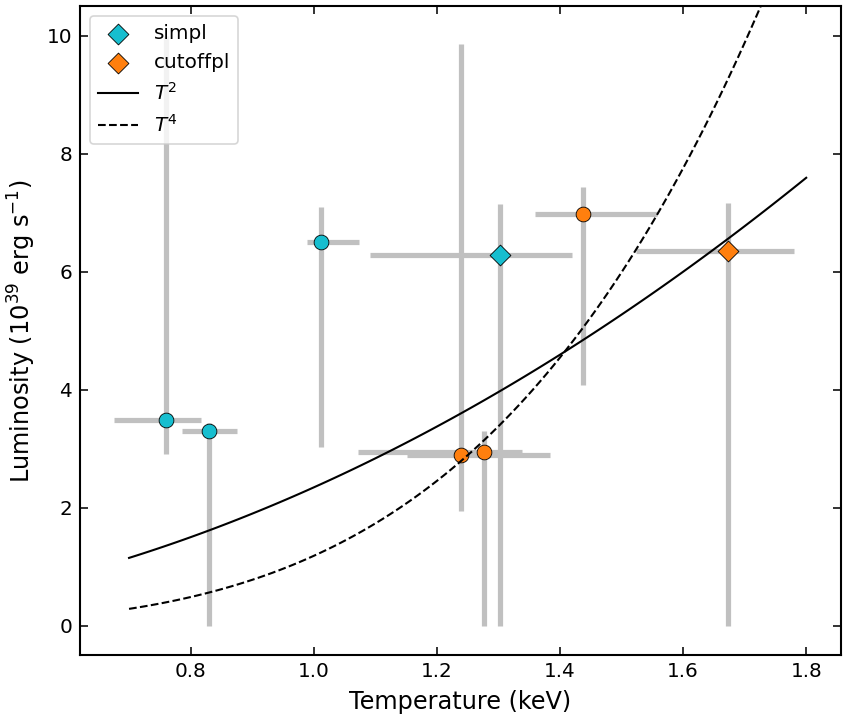}
   \caption{Luminosity-temperature relationship derived from \nicer+\nustar (diamond marker) and {\sl XMM-Newton} (circle marker) spectral fits. Unabsorbed luminosity was calculated assuming a distance of 2.9~Mpc \citep{2013AJ....146...86T}. {\bf Left panel}: {\sc diskbb} component. {\bf Right panel}: {\sc diskpbb} model. In both cases, $T^2$ and $T^4$ trends were added for reference.}
   \label{fig:templumin}
\end{figure*}

The case of a magnetic accretor, specifically a ULX pulsar, is based on the framework outlined in \citep{Walton2018ApJ...856..128W,Walton2018MNRAS.473.4360W}. Such a framework consists of two thermal blackbody components that represent the accretion flow emanating from the magnetosphere, the point where the NS's magnetic field cuts off the disk, causing the accreted material to stick to the magnetic field lines instead of falling directly onto the star. In addition, the central accretion columns formed by the material flow at the magnetic poles are described by a power-law component with an exponential cutoff ({\sc cutoffpl}).

For the thermal components, we again adopted the {\sc diskbb}+{\sc diskpbb} combination often used in ULX pulsar studies. The spectral shape of the accretion columns in the NGC 4190 ULX1 source cannot be determined directly, since no pulsations have been detected. 
To model the high-energy part of the spectrum ($E>10$~keV), we used a method similar to \citet{Walton2018MNRAS.473.4360W} by assigning the spectral parameters to the average values detected in the pulsed emission of the four ULX pulsars identified so far. These values include $\Gamma$ = 0.59 and $E_{\rm cut}$ = 7.9 keV for the {\sc cutoffpl} component \citep{Brightman2016ApJ...816...60B, Walton2018ApJ...856..128W}. 


The results of applying both phenomenological models to the \nicer+\nustar data are shown in \hyperref[tab:params]{Table~\ref{tab:params}} and \hyperref[fig:datafit]{Figure~\ref{fig:datafit}}. We obtain good fits overall in both cases, with reduced $\chi^2$ of ${\sim}$1.1. Both models show similar local absorption columns of $N_{\rm H}{\sim}0.1{\times}10^{22}$~cm$^{-2}$, cold {\sc diskbb} temperatures of ${\sim}$0.4 keV, and hot {\sc diskbb} temperatures higher than 1 keV, with $p$ indices greater tha 0.7.
The unabsorbed luminosity of NGC 4190 ULX1 yields ${\sim}7.6{\times}10^{39}$~erg~s$^{-1}$ for a distance of 2.9 Mpc in the $0.3-30$ keV energy range.

We used MCMC chains to compute the errors after generating chains of 10$^6$ samples using the {\rm XSPEC} {\sc chain} task with walkers equal to 16 times the number of free parameters, and checking that each parameter converged successfully (i.e., auto-correlation time close to unity; see \citealt{Fogantini2023, saavedragx13} for more details).

We also modeled archival XMM-Newton pn+MOS observations separately, using both proposed scenarios, in order to use the associated temperatures in the next section. The same XMM-Newton dataset was presented and analyzed, firstly by \cite{Ghosh2021}. As the EPIC background was significant above 10 keV in all observations, in order to properly account for the high-energy-emission components, we froze the {\sc simpl} parameters to those obtained using \nicer+\nustar data. We left only the {\sc simpl} and {\sc cutoffpl} normalization parameters free to vary. The parameters of each model are detailed in \hyperref[tab:paramsxmm]{Table~\ref{tab:paramsxmm}} in the appendix. 

\section{Discussion}

We examined the entire collection of archived X-ray observations from \ngc, with particular emphasis on the concurrent \nicer and \nustar observations. Our analysis included examination of the power spectra in the 0.01--10 Hz range, which showed no associated pulsations.  We placed an upper limit on the detectable PFs for \nicer and \nustar derived from simulations. We applied phenomenological models to fit the time averaged spectra, focusing on scenarios involving compact objects with and without magnetic fields, such as an NS or BH. 

To classify the source, we compared its spectral parameters with those of other sources and created a hardness–luminosity diagram. We used the energy range of 0.3--1.5 keV as the soft band and 1.5--10 keV as the hard band, following \citet{Gurpide2021A&Aa}. The ratio between the hard and soft bands was found to be $\sim$2.4, placing it in the same region as both pulsating and non-pulsating ULXs. Moreover, the flux ratio between the {\sc cutoffpl} component and the total 0.3--30 keV flux is ${\sim}28\%$, placing NGC~4190~ULX1 closer to non-pulsating sources \citep{Walton2018ApJ...856..128W}.
Additionally, we observed that the spectra are similar to that of NGC 5204 X-1 in an HUL state \citep{Gurpide2021A&Ab}.

We now focus on the temporal analysis of the behavior of the thermal components, examining their evolution in a plot of luminosity versus temperature. We calculated the unabsorbed luminosity of each thermal component over a wide energy range, from 0.3 to 30 keV. \hyperref[fig:templumin]{Figure~\ref{fig:templumin}} shows a comparison of temperature and luminosity for {\sc diskbb} and {\sc diskpbb} components. In each panel we have plotted the $T^2$ and $T^4$ trends. 


It is valuable to compare the observed trends with those projected in theoretical scenarios of the luminosity-temperature relation, such as a Shakura-Sunyaev sub-Eddington thin disk \citep[where $L$ is proportional to $T^4$ for a constant emitting area,][]{1973A&A....24..337S}, or a thick disk where advection is significant \citep[where $L$ is proportional to $T^2$,][]{2000PASJ...52..133W}. 
Based on the analyzed archive data from \nicer, \nustar, and {\sl XMM-Newton}, it is clear that  it favors a proportional relationship between $L$ and $T^2$, no matter which compact object scenario we consider. This implies the presence of advection in the system. 


We now present a more physical model consistent with the newly available data, based on the phenomenological fit of the X-ray observations. The model extends that presented by \cite{2023A&A...671A...9A}, which was based on \textit{XMM-Newton} data alone. A physically motivated model offers deeper insights into the possible mechanisms operating in the source and opens the possibility to new observational tests.

\subsection{Disk and wind}

We assumed that the accreting object is a BH of $10 M_{\odot}$. It forms a close binary with an evolved massive star. \cite{2013ApJS..206...14G} imposed constraints on the optical counterpart, suggesting that if $M_{\rm BH}=10 M_{\odot}$, the mass of the star should be $< 50 M_{\odot}$ and its radius $< 86 R_{\odot}$. These constraints are met by a star of type B2V, for example. 

We assumed that the BH accretes matter from the star at a rate of $10\dot{M}_{\rm Edd}$ (we will discuss this assumption when we present the results). We denoted this initial rate as $\dot{M}_{\rm input}$. Then, at some critical distance, $r_{\rm crit}$, from the BH, the radiation pressure of the disk exceeds the gravitational attraction of the compact object. Hydrodynamic equilibrium in the vertical direction is no longer maintained, and the disk begins to lose mass in the form of a wind. The critical radius is given by $r_{\rm crit} \sim 40\; \dot{m}_{\rm input} r_{\rm g}$, with $r_{\rm g}=GM_{\rm BH}/c^2$ and $\dot{m}_{\rm input}=\dot{M}_{\rm input} / \dot{M}_{\rm Edd}$ \citep{2004PASJ...56..569F}. The Eddington rate is
\begin{equation} \label{eq: tasa eddington}
    \Dot{M}_{\rm{Edd}}= \frac{L_{\rm{Edd}}}{\eta c^2} \approx 2.2\times 10^{-8} M_{\rm BH} \, {\rm yr^{-1}} = 1.4 \times 10^{18} \frac{M}{M_\odot} \, \rm{g \, s^{-1}},
\end{equation}
with $L_{\rm Edd}$ the Eddington luminosity (defined as the luminosity required to balance the attractive gravitational pull of the BH by radiation pressure), $\eta \approx 0.1$ the accretion efficiency, and $c$ the speed of light. 
The critical radius divides the disk into two regions: a standard outer disk that is optically thick but geometrically thin \citep{1973A&A....24..337S} and a radiation-dominated inner disk with advection \citep{2004PASJ...56..569F}. In this inner disk, the advection is parameterized as a fraction, $f$, of the viscous heating, $Q_{\rm adv}=fQ_{\rm vis}$, and the disk becomes geometrically thick. The ejection of winds by the radiation force helps to regulate the mass accretion rate onto the BH ($\dot{M}_{\rm acc}$) at the Eddington rate. We then have $\dot{M}_{\rm acc}=\dot{M}_{\rm input}$ in the outer region of the disk and $\dot{M}_{\rm acc}=\dot{M}_{\rm input} (r_{\rm d}/r_{\rm crit})$ in the inner region \citep{2004PASJ...56..569F} ($r_{\rm d}$ is the radial distance to the BH on the equatorial plane of the disk). The difference is the mass rate injected into the wind: $\dot{M}_{\rm w}=\dot{M}_{\rm input} - \dot{M}_{\rm acc} (r_{\rm d})$.

A self-similar solution for optically thick supercritical disks with mass loss can be found for the effective disk temperature, as in the appendix of \cite{2004PASJ...56..569F}. The solution reads

\begin{equation}\label{intensidad}
     \sigma T_{\rm eff}^4 =
    \left\lbrace \begin{array}{l}
   \dfrac{3GM_{\rm BH}\dot{M}_{\rm input}}{8\pi r_{\rm d}^3}{f_{\rm in}}, \ \ r_{\rm d} > r_{\rm crit}\\ \\
 \dfrac{3}{4}\sqrt{c_3}\dfrac{L_{\rm Edd}}{4\pi r_{\rm d}^2}, \ \ r_{\rm d} \le r_{\rm crit}
    \end{array} 
    \right.
,\end{equation}
with $\sqrt{c_3}=H/r_{\rm d}=\tan{\delta}$, where $H$ is the scale height of the disk, $\delta$ is the disk opening angle, and ${f_{\rm in}=1-r_{\rm in}/r_{\rm d}\approx1}$ \ (since ${r_{\rm d}>r_{\rm crit}}$, then ${r_{\rm d}\gg r_{\rm in})}$. Here, $r_{\rm in}$ is the location of the inner edge of the disk. The coefficient, $c_3$, depends on the advection parameter, the adiabatic index of the gas, $\gamma=5/3$, and the viscosity, $\alpha$ \citep[see Appendix in][]{2004PASJ...56..569F}. Here we assumed a disk with $f=0.5$ and $\alpha=0.5$; that is, we assumed equipartition between advection and viscous heating. The scale height of the disk (and thus its semi-opening angle) is completely determined by the coefficient, $c_3$, and therefore by $\gamma$, $f$, and $\alpha$.
The values assumed for these parameters lead to a disk opening angle of $\delta \sim 30^{\circ}$. It should also be noted that in the critical region of the disk $T_{\rm eff}\propto r^{-1/2}$, contrary to what is expected for a standard disk (see \citealp{2000PASJ...52..133W} for details).

The effects of the radiation on the matter located at the escape surface of the disk in the region, $r_{\rm d}\leq r_{\rm crit}$, have been calculated in detail by \cite{2023A&A...671A...9A}. The radiation field transfers both energy and angular momentum to the wind, which escapes the system with a strong equatorial component, forming a funnel above the BH. The wind itself is opaque until it reaches the photospheric radius, where it becomes transparent to its own radiation. This photosphere is defined by the condition that the optical depth, $\tau_{\rm photo}$, is unity for an observer at infinity. If the velocity of the wind is relativistic, the optical depth in the observer's frame depends in general on the magnitude of the velocity and the viewing angle. The position of the apparent photosphere from the equatorial plane, $z_{\rm photo}$, is given by \citep{2023arXiv231115050A}

\begin{equation}
    \tau_{\rm photo}=\int^\infty_{z_{\rm photo}} \gamma_{\rm dw}(1-\beta \cos{\theta}) \, \kappa_{\rm co} \,\rho_{\rm co} {\rm d}z =1
,\end{equation}
where $\gamma_{\rm dw}$ is the wind Lorentz factor, $\kappa_{\rm co}$ is the opacity in the comoving frame, and $\rho_{\rm co}$ is the comoving wind density. Since we assumed a fully ionized wind, the opacity is dominated by free electron scattering ($\kappa_{\rm co}=\sigma_{\rm T}/m_{\rm p}$). 

\begin{table} 
\begin{center}
\caption{Parameters of the model.}
\label{tab:parametros_generales}
\begin{adjustbox}{max width=\columnwidth}
\begin{tabular}{l c c c}
\hline
\hline
\rule{0pt}{2.5ex}Parameter & Symbol & Value & Units  \\
\hline
\rule{0pt}{2.5ex}Black hole mass$^{(1)}$ & $M_{\rm BH}$    & 10 & $M_{\odot}$ \\
Gravitational radius$^{(2)}$ & $r_{\rm g}$   & $1.48\times 10^6$  & $\rm{cm}$ \\
Critical radius$^{(2)}$ & $r_{\rm crit}$ & $3.5\times 10^9$ &  ${\rm cm}$ \\
Eddington accretion rate$^{(2)}$ & $\dot{M}_{\rm Edd}$ & $2.2\times 10^{-7}$ & $M_{\odot} \ \rm{yr}^{-1}$ \\
Mass accretion rate$^{(1)}$ & $\dot{M}_{\rm input}$ & $2.2\times 10^{-6}$ & $M_{\odot} \ \rm{yr}^{-1}$ \\
Hot gas Lorentz factor$^{(2)}$ & $\gamma_{\rm dw}$ & $1.25$ &  \\
Geometric beaming factor$^{(2)}$ & $b$ & 0.73 &  \\
Hadron-to-lepton ratio$^{(1)}$ & $a$ & $0.01$ &  \\
Kinetic power of electrons$^{(2)}$ & $L^{\rm dw}_{\rm K}$ & $2.2\times 10^{39}$ & ${\rm erg \ s^{-1}}$ \\
Content of relativistic particles$^{(1)}$ & $q_{\rm rel}$ & 0.15 &  \\
Injection spectral index$^{(1)}$ & $p$ & $2$ &  \\
Viewing angle$^{(1)}$ & $i$& $\approx 0$ & degrees\\
\hline
\end{tabular}
\end{adjustbox}
\end{center}
\footnotesize{\textbf{Notes.} We indicate the parameters that we have assumed with superscript ${(1)}$  and those that we have derived with ${(2)}$.}
\end{table}

\subsection{Emission}

When the system is viewed face-on (i.e., with an inclination angle of $i\approx 0^{\circ}$), the only X-ray emission visible from the disk is that which escapes through the funnel. This emission is produced near the innermost part. The radiation from the rest of the disk is absorbed and reprocessed in the wind and shifted to lower energies \citep[see][]{2023A&A...671A...9A}. 

The effect of geometrical beaming is to cause an observer to infer an isotropic luminosity given by
\begin{equation}
L_{\rm iso}\approx L_{\rm Edd}\left[ 1+ \frac{\ln{\dot{m}}}{b}\right],\label{eq:Liso}
\end{equation}
where $b$ is the beaming factor \citep{2008MNRAS.385L.113K}. This factor varies with the accretion rate (in Eddington
units) as $b\propto \dot{m}^{-2}$ \citep{2023NewAR..9601672K}\footnote{Note that all the emission from the disk reaching the observer is produced in its innermost region, as the remaining radiation from the outer part is absorbed by the dense wind. The beaming factor will then affect the disk spectrum uniformly on all energies.}; in our case, we obtain $b=0.73$. However, the emission extends to energies higher than 10 keV that cannot be produced in the disk; the hard X-rays found in our observations require an additional component. Such a component could be a hot plasma in the funnel above the BH. This plasma can be heated by magnetic reconnection, and a nonthermal particle population could be created by turbulent acceleration \citep{2016MNRAS.463.4331D}. The relativistic electrons in the plasma then Comptonize the softer photons from the disk, producing the nonthermal tail. 

We modeled the X-ray emission discussed in Sect. \ref{sec:results} using the model just described. The adopted and derived parameters are presented in \hyperref[tab:parametros_generales]{Table~\ref{tab:parametros_generales}}.  

\begin{figure}[h!]
  \centering
  \begin{minipage}{0.48\textwidth}
    \centering
    \includegraphics[width=\columnwidth]{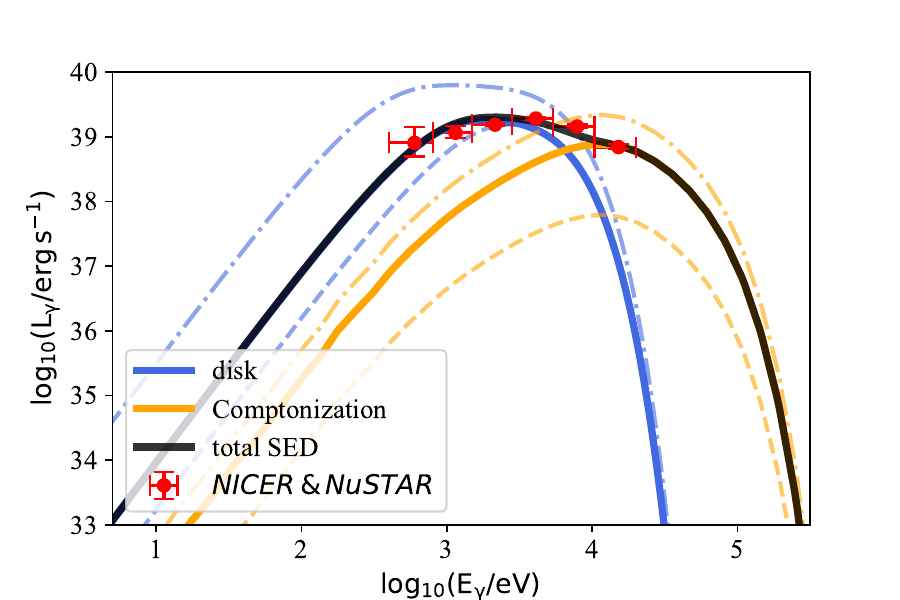}
  \end{minipage}
  \hfill
  \begin{minipage}{0.49\textwidth}
    \centering
    \includegraphics[width=\columnwidth, height=4.2cm]{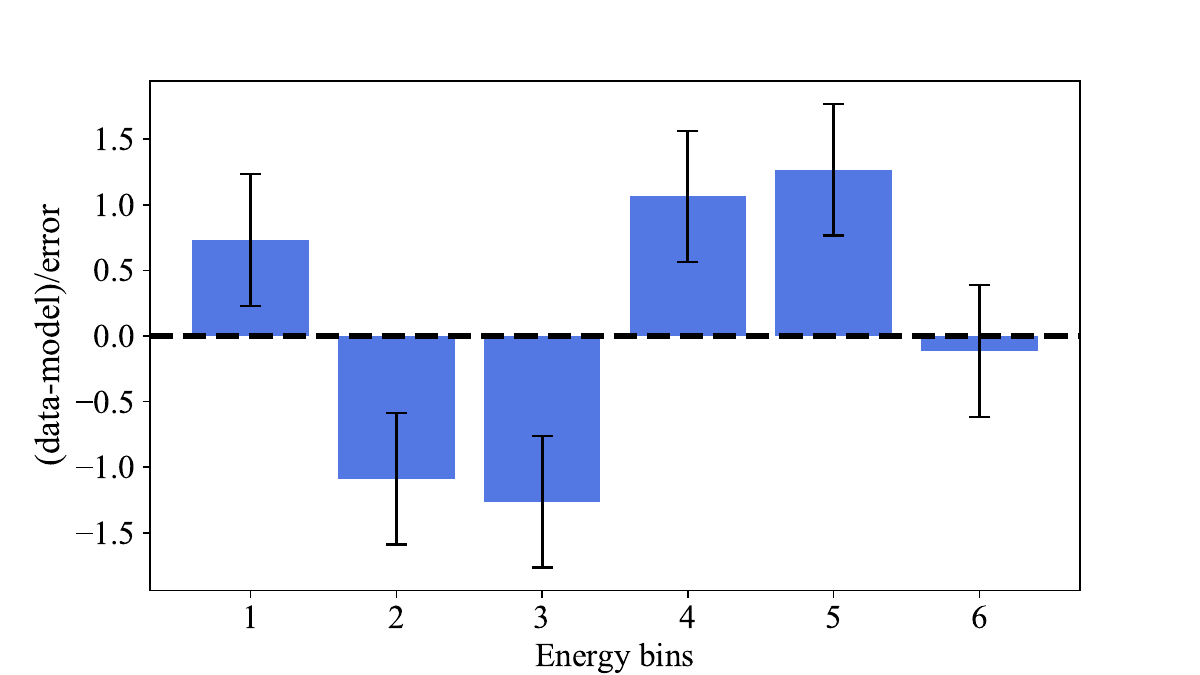}
  \end{minipage}
  \hfill
  \caption{{Top}: SED of the X-ray emission of NGC 4190 ULX-1 on a logarithmic scale. We plot the radiation from the innermost part of the disk that escapes through the funnel (blue line) and the radiation produced by Comptonization of the disk emission by nonthermal electrons in the funnel (orange line). The black line represents the total SED, while red dots correspond to energy-binned observations from \nicer and \textit{NuSTAR}. We also present, for comparison, the application of our model assuming an accretion rate of $ 1\,\dot{M}_{\rm Edd}$ (dashed lines) and of $15\,\dot{M}_{\rm Edd}$ (dash-dotted lines). \textit{Bottom:} Residuals of the best fit. The six spectral bins correspond to the data in the $0.4-20$ keV energy range.}
  \label{fig:total_sed}
\end{figure}

\begin{figure}
    \centering    \includegraphics[width=\columnwidth]{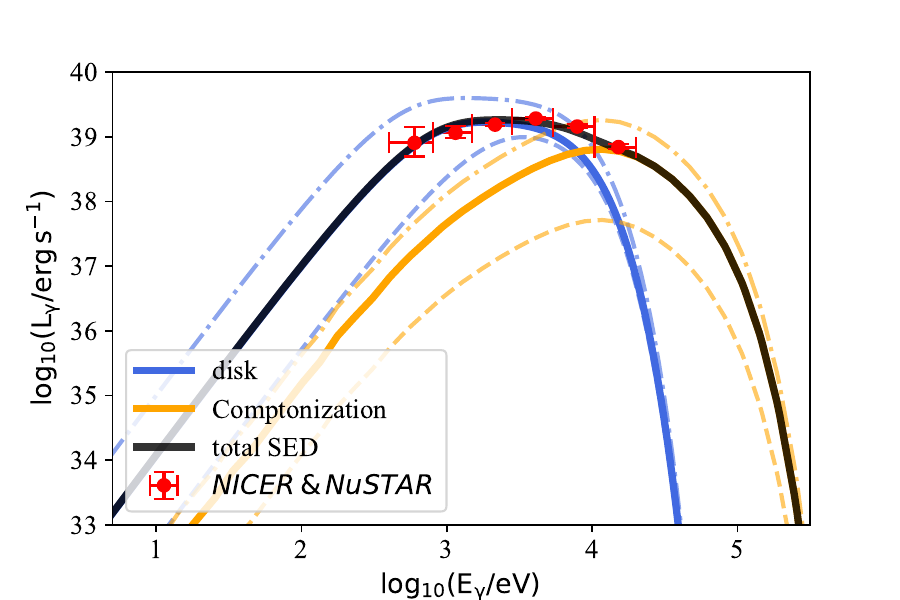}
    \caption{Alternative model: SED of the X-ray emission of NGC 4190 ULX-1 on a logarithmic scale for a different, less conservative, set of parameters. Here we fixed the BH mass to $5\,M_{\odot}$, the viscosity parameter to 0.1, and the content of relativistic particles to 0.25.  We also show for comparison the application of our model assuming an accretion rate of $ 1\,\dot{M}_{\rm Edd}$ (dashed lines) and of $15\,\dot{M}_{\rm Edd}$ (dash-dotted lines).}
    \label{fig:total_sed_B}
\end{figure}

The results are shown in the top panel of \hyperref[fig:total_sed]{Figure~\ref{fig:total_sed}}, where we plot the disk emission, along with the Comptonization produced by the nonthermal electrons in the funnel. To explore the space parameter that is physically allowed by the observation, we show our results for three different accretion rates: 1, 10, and 15 Eddington rates. 
As can be seen from the spectral energy distribution (SED), the model is in good agreement with the observational data for the set of parameters presented in \hyperref[tab:parametros_generales]{Table~\ref{tab:parametros_generales}}, extending the results of \cite{2023A&A...671A...9A}. We obtain a chi-square value of $\chi^2\sim 6.06$ for six spectral bins and show the residuals in the bottom panel of \hyperref[fig:total_sed]{Figure~\ref{fig:total_sed}}. An accretion rate higher than $10\dot{M}_{\rm Edd}$ flattens the spectrum, resulting in a hardening of the emission at soft X-rays, which mismatches the observational data at low energies. Additionally, the accretion rate influences two other parameters: the beaming factor and the available power for the hot gas, causing variations in the observed luminosity with $\dot{M}_{\rm input}$. In \hyperref[fig:total_sed_B]{Figure~\ref{fig:total_sed_B}} we present an alternative model with a smaller BH. The results are similar: essentially a lighter BH is possible, as long as the accretion rate remains at a level of ten Eddington rates. This result is natural, since the luminosity of the innermost region of the disk depends weakly on the normalized accretion rate (see Eq. \ref{eq:Liso}), and hence on the mass of the BH. However, this set of values is less conservative than the one in \hyperref[fig:total_sed]{Figure~\ref{fig:total_sed}}.

\subsection{Further remarks}

\cite{2005PASJ...57..691F} estimated the final velocity of a radiation-driven jet in a supercritically accreting BH. He found that the final Lorentz factor should be $\gamma_{\infty}\approx 1+ 0.223 \Gamma$, with $\Gamma= L/L_{\rm Edd}$, where $L$ is the disk luminosity. Since the wind regulates the disk luminosity to $L\sim L_{\rm Edd}$, one typically getst $\gamma_{\infty} \sim 1.2$ for the jet. This is essentially the same value that we obtained by considering the effect of the radiation force on a particle in the funnel, following the calculation method described in \cite{2023A&A...671A...9A}, see \hyperref[tab:parametros_generales]{Table~\ref{tab:parametros_generales}}.

Since the radiation pressure in the funnel above the BH exceeds the gravitational pull, the plasma cannot be at rest and must be constantly removed and replenished. Since heating the gas to Comptonization temperatures requires time for thermalization, we prefer to assume a nonthermal component flowing through the funnel. 
Whether this gas is considered a non-static corona or a mildly relativistic jet is more a matter of convention. \cite{2005ApJ...635.1203M} have shown that the observational features are quite similar in both cases. 

A collimated jet can manifest its effects at greater distances when it strikes the interstellar medium. There, a system of shocks may form. The forward shock is expected to be radiative and inefficient for particle acceleration, but the reverse shock could reaccelerate electrons to relativistic energies and produce radio synchrotron emission. This radiation could be detected by radio interferometric observations. If the jet has a hadronic content and the surrounding medium is dense enough, a $\gamma$-ray source could also be produced through the $p+p\rightarrow p+p+\pi^0$ channel, although it would be undetectable in the case of an extragalactic source like NGC 4190 ULX1. In galactic super-Eddington sources such as SS433, $\gamma$-rays have instead been reported by various instruments \citep[e.g.][]{2015ApJ...807L...8B}. Another potential super-Eddington microquasar with $\gamma$-ray emission seems to be 4FGL J1405.1-6119 \citep[see][]{saavedra}.

The conditions in the funnel are not expected to be very stable, due to various instabilities, clumping of the wind, and magnetic reconnection. Hence, some degree of rapid variability could be present, in accordance with some of the hints in this sense from our light curves. Longer observations will test this in real time. Differences with the results presented in the soft X-ray band presented by \cite{Ghosh2021} are already indicative of changes in the beaming.




A final note on ultraviolet (UV) radiation: the predicted emission in the UV range from the source is  $\sim10^{37}\,{\rm erg\,s^{-1}}$, which corresponds to the radiation from the inner disk. This is orders of magnitude lower than that of the ULX NGC 6946 X-1 investigated by \cite{2010ApJ...714L.167K}. Two possible explanations for the difference in the UV luminosity between NGC 4190 ULX1 and this source could be: i) the blackbody emission of the companion massive star in NGC6946 X-1, whose spectrum peaks in the UV range; and ii) the accretion rate of the compact object (higher $\dot{M}_{\rm input}$ leading to a spectrum flattening), resulting in increased luminosities at lower energies \citep[see e.g.,][]{2004PASJ...56..569F}.

Future high-resolution observations of NGC 4190 ULX1 with {\it Chandra} and VLBI may help to complete the picture of this source; in particular, its impact on the surrounding medium.


\section{Conclusions} \label{sec:concl}

In this paper we have performed a temporal and spectral study of the ultraluminous source NGC 4190 ULX1 using simultaneous \nustar and \nicer observations to understand the nature of the accretion disk around the BH, to detect spectral signatures of outflows, and to search for possible pulsations. The temporal analysis shows that the source is in a quiescent state, with a slight decrease in the count rate toward the end of the observation. No pulsations are detected, although spectral markers such as flux ratios do not allow us to discard the pulsating nature of the compact object. The source shows typical ULX behavior and its spectrum can be well fitted with two thermal blackbody components plus a Comptonization tail at high energies. The luminosity in the thermal region is consistent with an inflated radiation-dominated disk with advection. This type of disk loses mass due to strong winds. The wind is opaque, so the disk emission is only visible when the system is viewed nearly face-on, through the funnel above the BH. The Comptonization tail found in the \nustar data indicates the presence in this funnel of a population of electrons capable of upscattering softer photons from the inner disk region. We were able to model the data assuming a BH of 10 $M_{\odot}$ and an accretion rate of $\sim 10 \;\dot{M}_{\rm Edd}$. The matter escaping from the funnel could form a jet. Future high-resolution radio observations may reveal synchrotron radiation associated with the jet, while the detection of Balmer and He lines may provide more information about the wind, in particular its velocity.

\begin{acknowledgements}
  
We are grateful to the anonymous reviewer for their insightful comments and suggestions, which have helped us to improve the clarity and quality of our manuscript.
JAC, EAS, FAF and FG acknowledge support by PIP 0113 (CONICET). FAF is fellow of CONICET. JAC, GER, and FG are CONICET researchers. JAC is a Mar\'ia Zambrano researcher fellow funded by the European Union  -NextGenerationEU- (UJAR02MZ). This work received financial support from PICT-2017-2865 (ANPCyT). This work was partly supported by the Centre National d'Etudes Spatiales (CNES), and based on observations obtained with MINE: the Multi-wavelength INTEGRAL NEtwork. JAC was also supported by grant PID2019-105510GB-C32/AEI/10.13039/501100011033 from the Agencia Estatal de Investigaci\'on of the Spanish Ministerio de Ciencia, Innovaci\'on y Universidades, and by Consejer\'{\i}a de Econom\'{\i}a, Innovaci\'on, Ciencia y Empleo of Junta de Andaluc\'{\i}a as research group FQM-322, as well as FEDER funds. GER acknowledges financial support from the State Agency for Research of the Spanish Ministry of Science and Innovation under grants PID2019-105510GB-C31AEI/10.13039/501100011033/ and
PID2022-136828NB-C41/AEI/10.13039/501100011033/, and by “ERDF A way of making Europe”, by the "European Union", and through the "Unit of Excellence Mar\'ia de Maeztu 2020-2023" award to the Institute of Cosmos Sciences (CEX2019-000918-M). Additional support came from PIP 0554 (CONICET). FG acknowledges support from PIBAA 1275 (CONICET). 

\end{acknowledgements}

\bibliographystyle{aa}
\bibliography{biblio}

\appendix

\section{Additional table}

\hyperref[tab:paramsxmm]{Table A} shows in detail all the parameters of the models fitted to the XMM-Newton dataset. Luminosities are estimated in the 0.3-30 keV energy range, using a distance of 2900~kpc. Only the normalizations of high energy components were left free to vary. 

\renewcommand{\arraystretch}{1.5} 
\begin{table*}
\begin{center}
    
\caption{Best fit parameters and 90\% level uncertainties derived from {\sl XMM-Newton} average spectra.}
\label{tab:paramsxmm}

\begin{tabular}{ c  c  c  c  c  c  c  c }
    \hline
    Component           & Parameter                             & \multicolumn{6}{c}{ Observation } \\ 
                        &                                       & \multicolumn{2}{c}{0654650101} & \multicolumn{2}{c}{0654650201} & \multicolumn{2}{c}{0654650301} \\ 
                        &                                       & {\sc simpl}  & {\sc cutoffpl}  & {\sc simpl} & {\sc cutoffpl}  & {\sc simpl} & {\sc cutoffpl} \\ \hline
    
    {\sc cons}          & $C_{\rm MOS1}$                        & $0.99\pm0.03$ & $0.99\pm0.03$ & $1.09^{+0.02}_{-0.03}$ & $1.08^{+0.02}_{-0.03}$ & $1.01\pm0.02$ & $1.01^{+0.03}_{-0.02}$ \\
    
                        & $C_{\rm MOS2}$                        & $0.95\pm0.03$ & $0.94\pm0.03$ & $1.04^{+0.03}_{-0.02}$ & $1.04^{+0.02}_{-0.03}$ & $1.03\pm0.02$ & $1.03\pm0.02$ \\ 
                        
    {\sc tbabs}         & $N_{\rm H}$ (10$^{22}$~cm$^{-2}$)     & $0.06^{+0.05}_{-0.04}$ & $0.03^{+0.04}_{-0.02}$ & $0.13^{+0.04}_{-0.03}$ & $0.1\pm0.03$ & $0.09\pm0.03$ & $0.1^{+0.02}_{-0.03}$ \\ 
    
    {\sc diskbb}        & $kT_{\rm in}$ (keV)                   & $0.3\pm0.1$  & $0.4\pm0.1$  & $0.24\pm0.03$  & $0.29^{+0.05}_{-0.04}$ & $0.3^{+0.06}_{-0.05}$ & $0.3^{+0.08}_{-0.05}$ \\                  
   
    {\sc cutoffpl}      & $\Gamma$  & -- & $0.59^{\dagger}$ & -- & $0.59^{\dagger}$ & -- & $0.59^{\dagger}$ \\
     
                        & $E_{\rm fold}$ (keV)  & -- & $7.1^{\dagger}$  & -- & $7.1^{\dagger}$  & -- & $7.1^{\dagger}$ \\ 
    
    {\sc simpl}         & $\Gamma$  & $3.4^\dagger$ & -- & $3.4^\dagger$ & -- & $3.4^\dagger$ & --  \\
    
                        & $CF$ & $0.9^\dagger$ & -- & $0.9^\dagger$ & --  & $0.9^\dagger$ & --  \\
                        
    {\sc diskpbb}       & $kT$ (keV)                            & $0.84^{+0.1}_{-0.06}$  & $1.3^{+0.1}_{-0.2}$   & $0.76\pm0.03$  & $1.23^{+0.09}_{-0.06}$ & $1^{+0.1}_{-0.05}$  & $1.4\pm0.2$ \\ 
                        & $p$                                   & $0.9^{+0.1}_{-0.2}$    & $0.9^{+0.1}_{-0.2}$   & $0.95\pm0.05$  & $0.95^{+0.05}_{-0.2}$ & $0.9\pm0.1$ & $0.8^{+0.2}_{-0.1}$ \\
    Luminosity          & $10^{39}$~erg~s$^{-1}$                & \multicolumn{2}{c}{$3.7\pm0.4$}   & \multicolumn{2}{c}{$4.3\pm0.5$}  & \multicolumn{2}{c}{$7.5\pm0.7$} \\                    
    $\chi^2$/dof        &                                       & 299.3/232 & 297.8/231  & 244.1/232 & 222.0/231 & 263.7/274 & 264.3/273 \\ \hline
    
    \end{tabular}
\end{center}
\footnotesize{\textbf{Notes.} Unabsorbed luminosities are estimated in the 0.3-30 keV energy range, using a distance of 2900~kpc. The $\dagger$ symbol indicates the parameters was frozen during the fit.}
\end{table*}

\end{document}